\documentclass[aps,prl,twocolumn,showpacs,superscriptaddress,floatfix]{revtex4}

\usepackage{graphicx}
\usepackage{amsmath}
\usepackage{epsfig}
\usepackage{helvet}
\usepackage{amssymb}

\newcommand{\be}{\begin{equation}}
\newcommand{\ee}{\end{equation}}
\newcommand{\bea}{\begin{eqnarray}}
\newcommand{\eea}{\end{eqnarray}}

\newcommand{\ket}[1]{\mbox{$| #1 \rangle$}}

\begin{document}
\title{Frustrated antiferromagnets with entanglement renormalization:\\ ground state of the spin-$\frac{1}{2}$ Heisenberg model on a kagome lattice}
\author{G. Evenbly}
\author{G. Vidal}
\affiliation{School of Mathematics and Physics, the University of
Queensland, Brisbane 4072, Australia} 
\date{\today}

\begin{abstract}
Entanglement renormalization techniques are applied to numerically investigate the ground state of the spin-$\frac{1}{2}$ Heisenberg model on a kagome lattice. Lattices of $N=\{36,144,\infty\}$ sites with periodic boundary conditions are considered. For the infinite lattice, the best approximation to the ground state is found to be a valence bond crystal (VBC) with a 36-site unit cell, compatible with a previous proposal. Its energy per site, $E=-0.43221$, is an exact upper bound and is lower than the energy of any previous (gapped or algebraic) spin liquid candidate for the ground state.  
\end{abstract}


\maketitle

Low dimensional spin-$\frac{1}{2}$ quantum systems have long been the focus of intense research efforts, largely fueled by the search for exotic states of matter. An important example of a geometrically frustrated quantum antiferromagnet \cite{Review} is the spin-$\frac{1}{2}$ kagome-lattice Heisenberg model (KLHM). Despite a long history of study, the nature of its ground state remains an open question. Leading proposals include valence bond crystal (VBC) \cite{VBCold, VBCpin, VBCnew, VBCcore, Series, SeriesNew} and spin liquid (SL)  \cite{Sachdev, ExactDiag, Wang, Mila, Hastings, CritSpinLiq1, CritSpinLiq2, CritSpinLiq3, KagDMRG} ground states. Interest has been further stimulated by recent experimental work on Herbertsmithite ZnCu$_3$(OH)$_6$Cl$_2$, a possible physical realization of the model \cite{Herbert}.

Progress in our understanding of the KLHM has been hindered, as with many other models of frustrated antiferromagnets, by the inapplicability of quantum Monte Carlo methods due to the negative sign problem. Nevertheless, systems with up to 36 sites have been addressed with exact diagonalization \cite{ExactDiag,ExactDiag2}, whereas the density matrix renormalization group (DMRG) has been used to explore lattices of order $N\approx 100$ sites \cite{KagDMRG}. Unfortunately, it is very difficult to infer the nature of the ground state of an infinite system from these results. The reason is that these lattices are still relatively small given the 36-site unit cell of the leading VBC proposal, or the algebraic decay of correlations in some SL proposals. In larger systems, support for a SL ground state has also been obtained with a SL ansatz \cite{CritSpinLiq1, CritSpinLiq2, CritSpinLiq3}, whereas evidence for a VBC has been obtained for an infinite lattice with a series expansion around an arbitrary dimer covering \cite{Series,SeriesNew}, but both approaches are clearly biased.

%%%%%%%%%%%%%%%%%%%%%%%%%%%%%%%%%%%%%%%
%%%%%%%%%%%%%%%%%%%%%%%%%%%%%%%%%%%%%%%
\begin{figure}[!htb]
\begin{center}
\includegraphics[width=8cm]{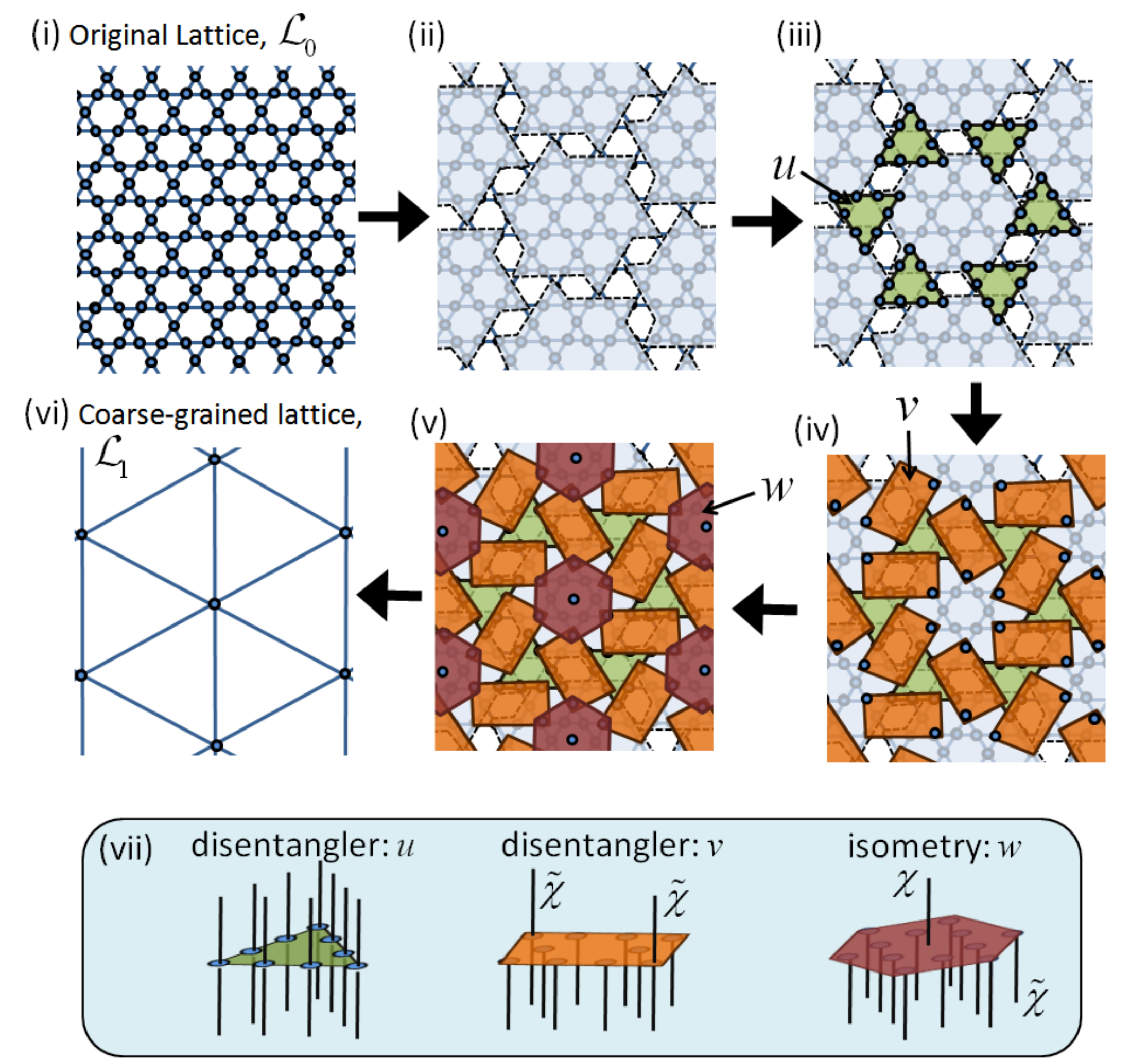}
\caption{(Color on-line) Coarse-graining transformation that maps (i) a kagome lattice $\mathcal L_0$ of $N$ sites into (vi) a coarser lattice $\mathcal L_1$ of $N/36$ sites. (ii) The lattice is first partitioned into blocks of 36 sites. (iii) Disentanglers $u$ are applied across the corners of three blocks, followed by (iv) disentanglers $v$ applied across the sides of two neighboring blocks. (v) Isometries $w$ map blocks to an effective site of the coarse-grained lattice. (vii) Tensors $u,v$ and $w$ have a varying number of incoming and outgoing indices according to Eq. \ref{eq:tensors1}. } \label{fig:KagScheme}
\end{center}
\end{figure}
%%%%%%%%%%%%%%%%%%%%%%%%%%%%%%%%%%%%%%%
%%%%%%%%%%%%%%%%%%%%%%%%%%%%%%%%%%%%%%

In this paper we report new, independent numerical evidence in favor of a VBC ground state for the KLHM model. This is done with entanglement renormalization \cite{ER, MERA, NewAlgorithm, FreeFermions, FreeBosons, Finite2D, ER2D}, a real space RG approach that, through the proper removal of short-range entanglement, is capable of providing an approximation to ground states of large 2D lattices \cite{ER2D} by means of the multi-scale entanglement renormalization ansatz (MERA) \cite{MERA}. After describing a scheme for the Kagome lattice with periodic boundary conditions, we address lattices of $N=\{36,144,\infty\}$ sites. Our simulations converge to a VBC state compatible with that first proposed by Marston and Zeng \cite{VBCold} and revisited by Nikolic and Senthil \cite{VBCnew}, and by Singh and Huse \cite{Series, SeriesNew}. For an infinite lattice we obtain an energy per site $E=-0.43221$. This energy corresponds to an explicit (MERA) wave-function and therefore provides us with a strict upper bound for the true ground state energy. Importantly, its value is lower than the energy of any existing SL ansatz on a sufficiently large lattice, which we interpret as strong evidence for a VBC ground state in the thermodynamic limit. Our results are also the first demonstration of the utility of entanglement renormalization to study 2D lattice models that are beyond the reach of quantum Monte Carlo techniques.

The present approach is based on the coarse-graining transformation of Fig. \ref{fig:KagScheme}, which is applied to a kagome lattice $\mathcal{L}_0$ made of $N$ sites. It maps blocks of 36 sites of $\mathcal{L}_0$ onto single sites of a coarser lattice $\mathcal{L}_1$ made of $N/36$ sites. A Hamiltonian $H_0$ defined on lattice $\mathcal{L}_0$ becomes an effective Hamiltonian $H_1$ on lattice $\mathcal{L}_1$. Analogously, the ground state $\ket{\Psi_0}$ of $H_0$ is transformed into the ground state $\ket{\Psi_1}$ of $H_1$. The transformation decomposes into three steps. Firstly disentanglers $u$, unitary tensors that act on $9$ sites, are applied across the corners of three neighboring blocks. Then disentanglers $v$ are applied across the sides of two neighboring blocks; these tensors reduce ten sites (each described by a vector space $\mathbb{C}_2$ of dimension $2$) into two effective sites (each described by a vector space $\mathbb{C}_{\tilde{\chi}}$ of dimension $\tilde{\chi}$). Finally isometries $w$ map the remaining sites of each block into a single effective site of $\mathcal L_1$. Thus the tensors $u$, $v$ and $w$,
\begin{eqnarray}
	u^{\dagger}\!&:& {\mathbb{C}_2}^{\otimes 9} \rightarrow {\mathbb{C}_2}^{\otimes 9}, ~~~~~~~~~~~~ u^{\dagger} u = I_{2^9},\nonumber\\
	v^{\dagger}\!&:& {\mathbb{C}_2}^{\otimes 10} \rightarrow {\mathbb{C}_{\tilde{\chi}}}^{\otimes 2}, ~~~~~~~~~~~ v^{\dagger} v = I_{\tilde{\chi}^2},\nonumber\\
  w^{\dagger}\!&:& {\mathbb{C}_2}^{\otimes 6} \otimes {\mathbb{C}_{\tilde{\chi}}}^{\otimes 6} \rightarrow {\mathbb{C}_{\chi}}, ~~~~w^{\dagger} w = I_{\chi}, \label{eq:tensors1}
\end{eqnarray}
transform the ground state $\ket{\Psi_0}$ of lattice $\mathcal{L}_0$ into the ground state $\ket{\Psi_1}$ of lattice $\mathcal{L}_1$ through the sequence 
\begin{equation}
	\ket{\Psi_0} \stackrel{u}{\rightarrow} \ket{\Psi_0'} \stackrel{v}{\rightarrow} \ket{\Psi_0''} \stackrel{w}{\rightarrow} \ket{\Psi_1}.
	\label{eq:Psi}
\end{equation}
The \emph{disentanglers} $u$ and $v$ aim at removing short-range entanglement across the boundaries of the blocks; therefore states $\ket{\Psi_0'}$ and $\ket{\Psi_0''}$ possess decreasing amounts of short range entanglement. If state $\ket{\Psi_0}$ only has short-range entanglement to begin with, then it is conceivable that the state $\ket{\Psi_1}$ has no entanglement left at all. For a finite lattice ($N=144$) we consider a state $\ket{\Psi_0}$ that, after the coarse-graining transformation, gives rise to an entangled state $\ket{\Psi_1}$ on $N/36=4$ sites. For an infinite lattice we make instead an important assumption, namely that $\ket{\Psi_1}$ is a product (non-entangled) state (this is equivalent to setting $\chi=1$ in Eq. \ref{eq:tensors1}).
How short-ranged must the entanglement in $\ket{\Psi_0}$ be for this assumption to be valid? By reversing the transformation on a product state $\ket{\Psi_1}$, it can be seen that each site in $\ket{\Psi_0}$ is still entangled with \emph{at least} $84$ neighboring sites.

%%%%%%%%%%%%%%%%%%%%%%%%%%%%%%%%%%%%%%%
%%%%%%%%%%%%%%%%%%%%%%%%%%%%%%%%%%%%%%%
\begin{figure}[!tb]
\begin{center}
\includegraphics[width=7.5cm]{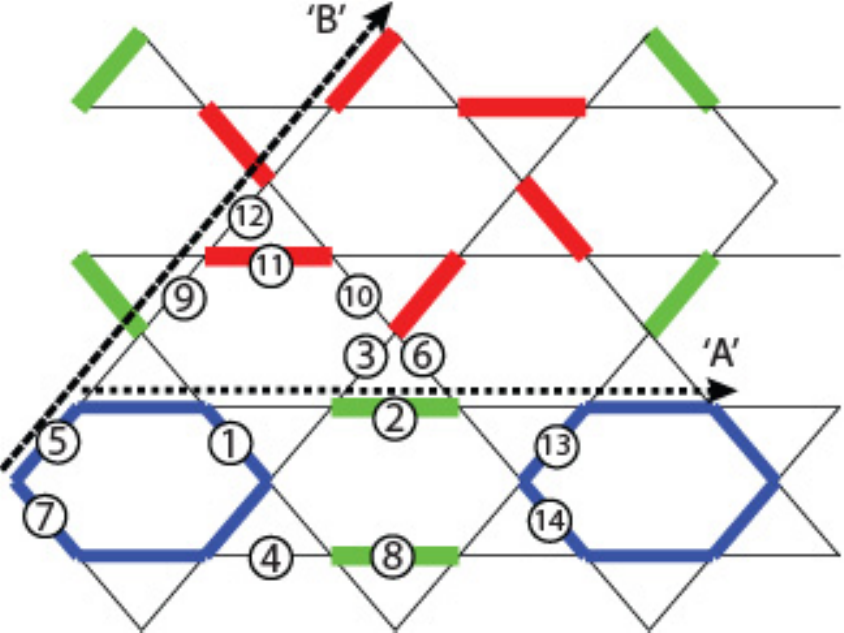}
\caption{(Color on-line) The 36-site unit cell for the honeycomb VBC, strong bonds are drawn with thick lines. Three different types of strong bonds can be identified; the six bonds belonging to the pinwheels (red), six bonds belonging to each `perfect hexagon' (blue) and the parallel bonds between perfect hexagons (green). Dotted arrows indicate the axis where spin-spin correlators have been computed. Bond-bond correlators have been computed between the reference bond (1) and the other numbered bonds.} \label{fig:UnitCell}
\end{center}
\end{figure}
%%%%%%%%%%%%%%%%%%%%%%%%%%%%%%%%%%%%%%%
%%%%%%%%%%%%%%%%%%%%%%%%%%%%%%%%%%%%%%%

The disentanglers and isometries $(u,v,w)$ were initialized randomly and then optimized so as to minimize the expectation value of the KLHM Hamiltonian,
\begin{equation}
H_0 = J\sum\limits_{\left\langle {i,j} \right\rangle } {{\rm{S}}_i  \cdot {\rm{S}}_j },
\label{eq:H}
\end{equation}
by following the algorithms of Ref. \cite{NewAlgorithm}, with cost $O(2^{12}\tilde \chi^6 \chi^2)$ \footnote[1]{The computational cost of simulations can be signifcantly reduced by incorporating symmetries of the system into the MERA tensor network \cite{Sukhi}. In this work, a U(1) symmetry is incorporated into the MERA to allow large $\tilde \chi$ simulations that would otherwise be unaffordable.}. Specifically, for lattices with $N=36$ and $144$ sites, the resulting (one-site and four-site) Hamiltonian $H_1$ is diagonalized exactly. Instead, for $N=\infty$, we use the finite correlation range algorithm (Sect. V.D of ref. \cite{NewAlgorithm}). All computations led to highly dimerized wave-functions of the VBC type. In order to explain the results, consider the \emph{exact} `honeycomb' VBC state, denoted $\ket{\textrm{h-VBC}}$, whose 36-site unit cell is shown in Fig. \ref{fig:UnitCell}. Each unit cell contains two `perfect hexagons' (resonating bonds around a hexagon) and a `pinwheel'. Three different types of strong bonds can be identified: those of the pinwheels (red), parallel bonds (green) and perfect hexagons (blue). The pinwheel and parallel bonds are singlets (energy per bond $=-0.75$) while the perfect hexagons are in the ground state of a periodic Heisenberg chain of 6 sites (energy per bond $=-0.4671$). The rest of links have zero energy. We call a `honeycomb' VBC a state that has strong bonds according to the above pattern, even though the rest of bonds (weak bonds) need not have zero energy. The `honeycomb' VBC was originally proposed by Marston and Zeng \cite{VBCold} (see also \cite{VBCnew,Series,SeriesNew}). Our simulations with $N=144$ and $N=\infty$ produce a VBC of this type as the best MERA approximation to the ground state.

%%%%%%%%%%%%%%%%%%%%%%%%%%%%%%%%%%%%%%%
%%%%%%%%%%%%%%%%%%%%%%%%%%%%%%%%%%%%%%%
\begin{table}[!b]
\caption{Ground state energies as a function of $\tilde\chi$.}
\begin{tabular}{cccc}
\hline\hline
 $\;\;\;\tilde\chi\;\;\;$  &   $\;\;N=\infty\;\;\;\;$   & $N=36$          & $N=36$        \\
         &                & (rand init)   & ($\ket{\textrm{h-VBC}}$ init)\\
   2     &    -0.42145    & -0.42164        & -0.42143      \\
   4     &    -0.42952    & -0.42816        & -0.42715      \\
   8     &    -0.43081    & -0.43199        & -0.43148      \\
   12    &    -0.43114    & -0.43371        & -0.43298      \\
   16    &    -0.43135    & -0.43490        & -0.43420      \\
   20    &    -0.43162    & -0.43611        & -0.43541      \\
   26    &    -0.43193    &                 &               \\
   32    &    -0.43221    &                 &               \\
\hline\hline
\end{tabular}
\label{table:Energy}
\end{table}
%%%%%%%%%%%%%%%%%%%%%%%%%%%%%%%%%%%%%%%
%%%%%%%%%%%%%%%%%%%%%%%%%%%%%%%%%%%%%%%

The energies obtained for an infinite lattice are shown in Table \ref{table:Energy}. For each value of $\tilde{\chi}$ (see Eq. \ref{eq:tensors1}), the MERA is an explicit wave-function and therefore provides an upper bound to the exact ground state energy. Energies computed for the $N=144$ lattice matched those of the infinite lattice to within $0.02\%$ and have been omitted. Our $N=\infty$ energy is also compatible with the energy $E=-0.433\pm0.001$ for a VBC obtained with series expansion in Ref. \cite{Series}, and is lower than that obtained in Ref. \cite{KagDMRG} with DMRG ($E = -0.43160$ for $N=108$) and in Ref. \cite{CritSpinLiq2} with fermionic mean-field theory and Gutzwiller projection ($E = -0.42863$ for $N=432$). We further notice that, where finite size effects are still relevant, such as in the $N=108$ case, they tend to decrease the ground state energy.

%%%%%%%%%%%%%%%%%%%%%%%%%%%%%%%%%%%%%%%
%%%%%%%%%%%%%%%%%%%%%%%%%%%%%%%%%%%%%%%
\begin{figure}[!tb]
\begin{center}
\includegraphics[width=8cm]{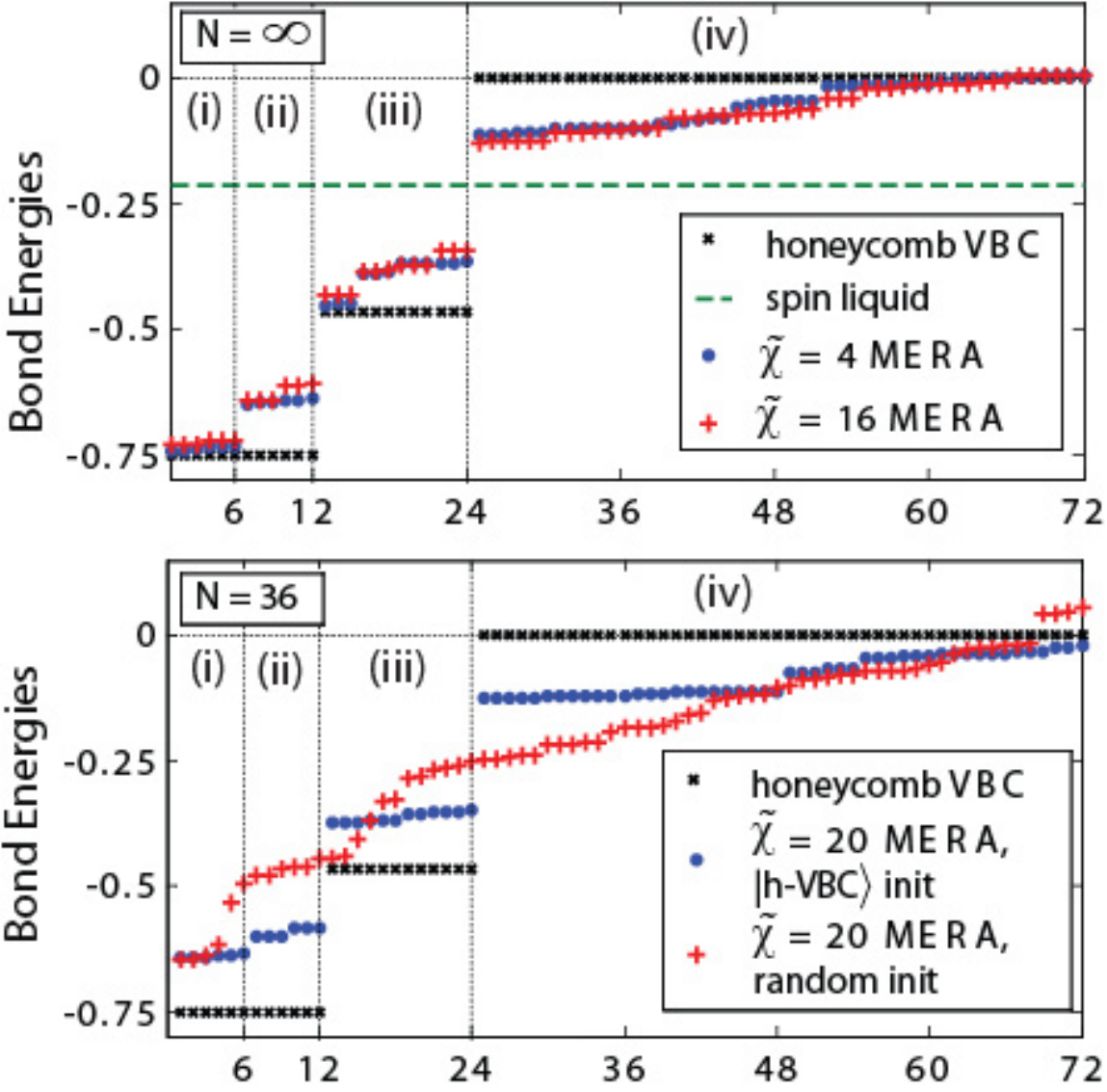}
\caption{(top) Bond energies for the 36-site unit cell of infinite MERA wave-functions, for two different values of $\tilde\chi$, as compared to those of an exact honeycomb VBC, $\ket{\textrm{h-VBC}}$, and those of a spin liquid, which by definition has all equal strength bonds. The MERA wave-functions clearly match the proposed honeycomb VBC; we identify (i) the six strong `pinwheel' bonds (red bonds), the six `parallel' bonds (green bonds) and (iii) the 12 `perfect hexagon' bonds (blue bonds). The (iv) remaining 48 bonds are the weak bonds of the unit cell. (bottom) Bond energies for the 36-site lattice. Here a randomly initialized MERA converges to a dimerized state that does not match the honeycomb VBC pattern, but gives lower overall energy than a honeycomb VBC initialized MERA of the same $\tilde \chi$.} \label{fig:BondEnergy}
\end{center}
\end{figure}
%%%%%%%%%%%%%%%%%%%%%%%%%%%%%%%%%%%%%%%
%%%%%%%%%%%%%%%%%%%%%%%%%%%%%%%%%%%%%%%

Fig. \ref{fig:BondEnergy} shows the distribution of bond energies obtained for the $N=\infty$ lattice. With $\tilde{\chi}=4$, one observes an energy increase per site over $\ket{\textrm{h-VBC}}$ of $\approx 0.08$ in the parallel (green) bonds and also in some of the hexagon (blue) bonds, with the weak bonds having lower energy in return. As $\tilde{\chi}$ is increased, the energy of the `strong' bonds becomes slightly larger and that of the `weak' bonds continues to decrease. However, the dimerization clearly survives: the bond energies are not seen to converge to a uniform distribution as required for a SL. 

%%%%%%%%%%%%%%%%%%%%%%%%%%%%%%%%%%%%%%%
%%%%%%%%%%%%%%%%%%%%%%%%%%%%%%%%%%%%%%%
\begin{figure}[!tb]
\begin{center}
\includegraphics[width=7cm]{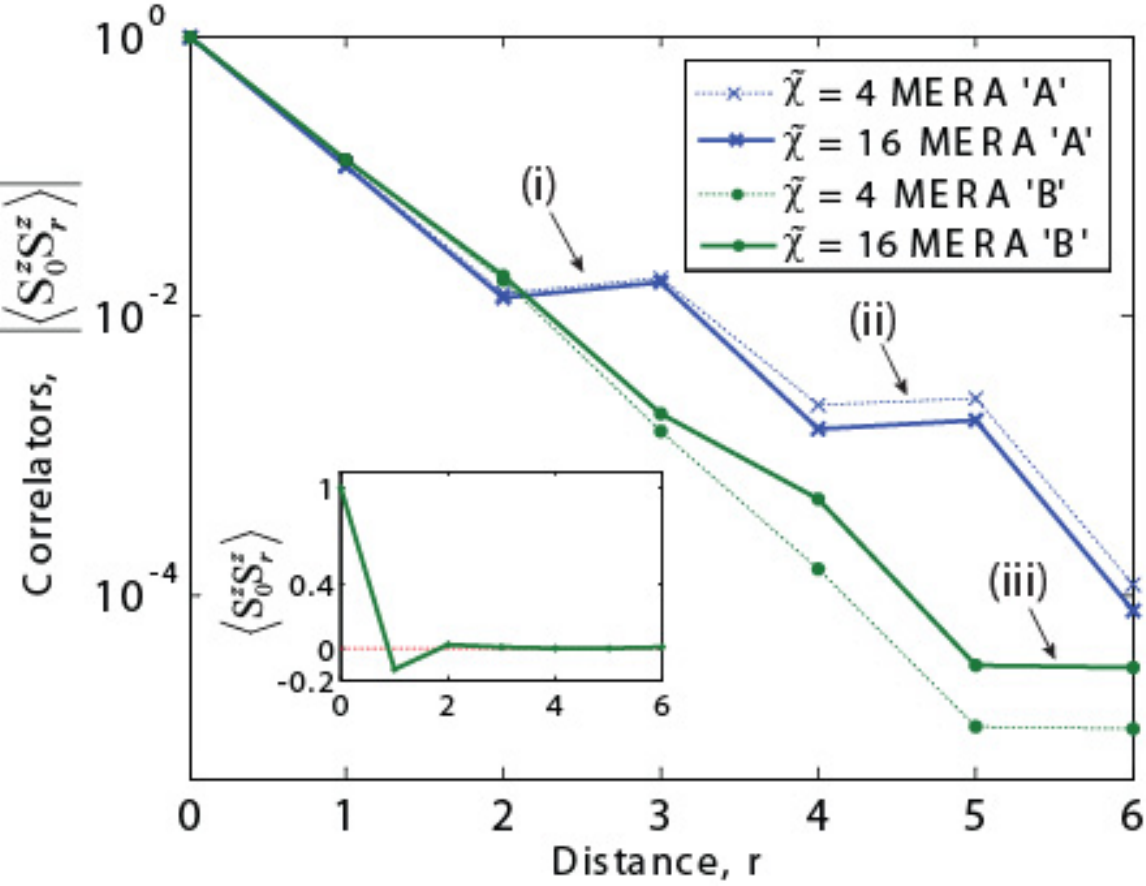}
\caption{Spin-Spin correlators along arrows `A' and `B' of Fig. \ref{fig:UnitCell} for infinite lattice MERA of $\tilde\chi=4$ and $\tilde\chi=16$. Although along both lattice directions considered the correlators decay exponentially, the decay along arrow `A' (a line joining two perfect hexagons) is seen to be slower than along arrow `B' (a line joining perfect hexagon to pinwheel). The plateaus marked (i), (ii) and (iii) show the correlation is the same with both spins of a strong bond.  } \label{fig:SpinSpinCorr}
\end{center}
\end{figure}
%%%%%%%%%%%%%%%%%%%%%%%%%%%%%%%%%%%%%%%
%%%%%%%%%%%%%%%%%%%%%%%%%%%%%%%%%%%%%%%

Fig. \ref{fig:SpinSpinCorr} shows spin-spin correlators evaluated along two different lattice axis $A$ and $B$ (cf. Fig. \ref{fig:UnitCell}) for $N=\infty$. These correlators decay exponentially with well defined `plateaus', where the correlation is the same with both spins of a strong bond. Correlations along the line joining a perfect hexagon and a pinwheel are seen to decay faster than along the line joining two perfect hexagons, consistent with the observation from Fig. \ref{fig:BondEnergy} that the pinwheel bonds remain almost exact singlets even for high values of $\tilde \chi$. Table \ref{table:BondCorr} shows bond-bond connected and disconnected correlators, $C_{1,\alpha}$ and $D_{1,\alpha }$, 
\begin{eqnarray}
C_{1,\alpha }  &\equiv& \left\langle {\left( {\vec S \cdot \vec S} \right)_1 \left( {\vec S \cdot \vec S} \right)_\alpha  } \right\rangle \\
D_{1,\alpha }  &\equiv&  C_{1,\alpha }  - \left\langle {\left( {\vec S \cdot \vec S} \right)_1 } \right\rangle \left\langle {\left( {\vec S \cdot \vec S} \right)_\alpha  } \right\rangle, \label{eq:Corr}
\end{eqnarray}
between a reference bond `1' and a surrounding bond $\alpha=1,\cdots, 14$ (cf. Fig.\ref {fig:UnitCell}). While disconnected correlators decay exponentially with distance, some connected correlators remain significant at arbitrary distances, demonstrating the long-range order of the VBC state. 

\begin{table}[!thb]
\caption{Bond-Bond correlators for  $\tilde\chi = 16$ ($N=\infty$).}
\begin{tabular}{cccc}
\hline\hline
Bond    & $C_{1,\alpha}$ & $D_{1,\alpha}$ & $\left\langle {\left( {\vec S \cdot \vec S} \right)_\alpha  } \right\rangle$ \\ 
   1      &  0.38639  &   0.22815         & -0.39780       \\
   2      &  0.29490  &   0.03496         & -0.65342       \\
   3      &  0.00422  &  -0.00115         & -0.01349       \\
   4      &  0.09503  &   0.04644         & -0.12211       \\
   5      &  0.22488  &   0.06609         & -0.39918       \\
   6      & -0.00609  &  -0.01426         & -0.02054       \\
   7      & -0.10372  &  -0.03376         & -0.34561       \\
   8      &  0.26083  &   0.01252         & -0.62421       \\
   9      &  0.00336  &   0.00368         &  0.00082       \\
   10     & -0.00001  &  -0.00017         & -0.00042       \\
   11     &  0.29113  &   0.00196         & -0.72693       \\
   12     &  0.00090  &  -0.00036         & -0.00321       \\
   13     &  0.16632  &   0.00026         & -0.41744       \\
   14     &  0.15674  &  -0.00037         & -0.39496       \\
\hline

\hline\hline
\end{tabular}
\label{table:BondCorr}
\end{table}

Let us discuss the results for a lattice with $N=36$ sites. When initialized with random tensors, the MERA produced VBC type configurations which typically did not match the honeycomb VBC, although simulations initialized in the state $\ket{\textrm{h-VBC}}$ retained the honeycomb VBC pattern (see Fig. \ref{fig:BondEnergy}). Here the randomly initialized VBC produced a lower energy ($0.5\%$ above the exact diagonalization result $E=-0.438377$ of \cite{ExactDiag}) than the honeycomb VBC type solution for an equivalent value of $\tilde{\chi}$ (cf. Table \ref{table:Energy}). These results strongly suggest that finite size effects in the $N=36$ site lattice lead to a significant departure from the physics of the infinite system. 

To summarize, we have used entanglement renormalization techniques to obtain new numerical evidence indicating that the ground state of the KLHM is of the honeycomb VBC type. In order to assess the robustness of this result, we briefly discuss some of the limitations of the present approach.

Firstly, the coarse-graining transformation of Fig. \ref{fig:KagScheme}, which maps 36 sites into one site, was designed to ensure compatibility with the 36-site unit cell of honeycomb VBC type solutions. While our approach did not preclude solutions with a smaller, compatible unit cell (such as a 12-site unit cell or a fully translation invariant solution), we cannot rule out the possibility that a state with an incompatible unit cell might have a lower energy. 

Secondly, the infinite lattice was investigated by restricting the range of entanglement in the ansatz to blocks of 84 spins, imposed through an unentangled state $\ket{\Psi_1}$ in Eq. \ref{eq:Psi}. This restriction was only implemented after preliminary simulations with $\tilde{\chi}=12$ had produced identical energies irrespectively of whether $\ket{\Psi_1}$ was allowed to be entangled. However, it could still be that entanglement in $\ket{\Psi_1}$ would make a big difference for larger values of $\tilde{\chi}$. We find this scenario quite unlikely, but could not test it due to computational limitations.

Finally, the MERA is an essentially unbiased method provided that the candidates to be the ground state of the system have all a relatively small amount of entanglement. But when deciding between a VBC (which mostly has short-range entanglement) and e.g. the algebraic SL of Refs. \cite{CritSpinLiq2,CritSpinLiq3} (significantly more entangled at all length scales), it might well be that the MERA is biased toward the low entanglement solution. Therefore our results do not conclusively exclude a SL ground state. We emphasize, however, that the ground state energies obtained with the MERA are lower than the SL energies of Refs. \cite{CritSpinLiq2, CritSpinLiq3, KagDMRG}.

We thank P. Lee, R. Singh, X.-G. Wen and S. White for valuable conversations. Support from the Australian Research Council (APA, FF0668731, DP0878830) is acknowledged.

\newpage

\textbf{Appendix A: The Triangular Lattice}

In this appendix we describe a coarse-graining transformation for the triangular lattice that, when used in conjunction with the scheme of Fig. \ref{fig:KagScheme}, allows one to analyze arbitrarily large or infinite kagome lattices without restricting $\ket{\Psi_1}$ in Eq. \ref{eq:Psi} to be unentangled. This scheme may also be used directly to simulate models defined on a triangular lattice.

%%%%%%%%%%%%%%%%%%%%%%%%%%%%%%%%%%%%%%%
%%%%%%%%%%%%%%%%%%%%%%%%%%%%%%%%%%%%%%%
\begin{figure}[!htb]
\begin{center}
\includegraphics[width=8cm]{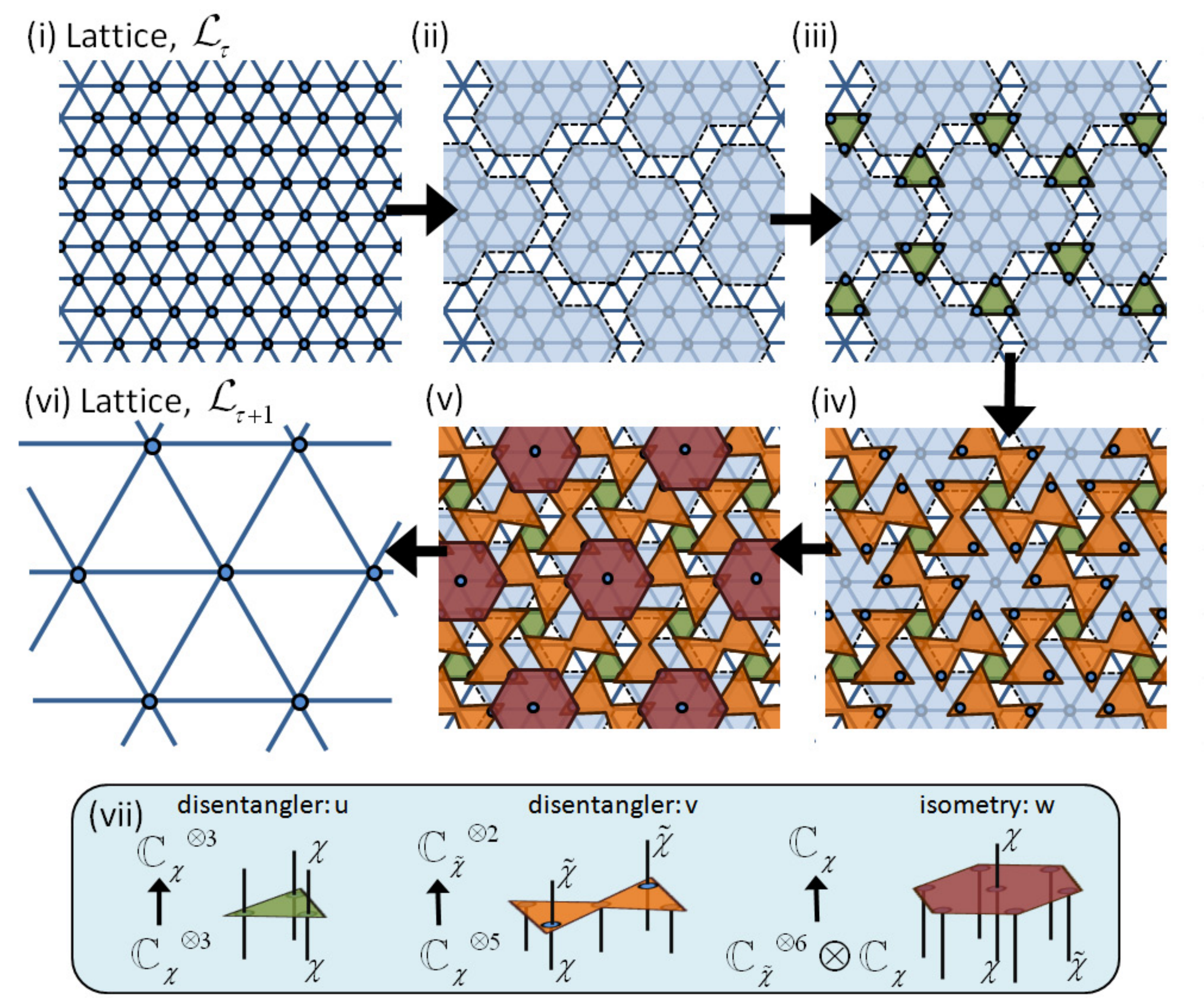}
\caption{(Color on-line) A coarse-graining scheme which transforms (i) a triangular lattice $\mathcal L_\tau$ of $N$ sites into (vi) a coarser triangular lattice $\mathcal L_{\tau+1}$ of $N/16$ sites. (ii) Firstly the lattice is partitioned into blocks of 16 sites. (iii) Disentanglers $u$ are applied across the corners of three blocks, followed by (iv) disentanglers $v$ applied across the sides of two neighboring blocks. (v) Isometries $w$ are applied to map blocks to an effective site of the coarse-grained lattice. (vii) Tensors $u,v$ and $w$ have a varying number of incoming and outgoing indices according to Eq. \ref{eq:tensors2}.}
\label{fig:TriScheme} 
\end{center}
\end{figure}
%%%%%%%%%%%%%%%%%%%%%%%%%%%%%%%%%%%%%%%
%%%%%%%%%%%%%%%%%%%%%%%%%%%%%%%%%%%%%%%

Fig. \ref{fig:TriScheme} shows a coarse-graining transformation that maps a triangular lattice $\mathcal L_\tau$ of $N$ sites into and a coarser triangular lattice $\mathcal L_{\tau+1}$ of $N/16$ sites. This transformation is accomplished in three steps; firstly disentanglers $u$, unitary tensors which act on three sites, are applied across the corner of three blocks. Next disentanglers $v$ are applied across the sides of neighboring blocks; these are isometric tensors that truncate five lattice sites (each one described by a vector space $\mathbb{C}_\chi$ of dimension $\chi$) into two effective sites (each one described by a vector space $\mathbb{C}_{\tilde\chi}$ of dimension $\tilde \chi$). Finally, isometries $w$ map the remaining sites of each block into a single effective site of $\mathcal L_{\tau+1}$. Collectively, the (isometric) tensors $u$, $v$ and $w$,
\begin{eqnarray}
	u^{\dagger}\!&:& {\mathbb{C}_\chi}^{\otimes 3} \rightarrow {\mathbb{C}_\chi}^{\otimes 3}, ~~~~~~~~~~~~ u^{\dagger} u = I_{\chi^3},\nonumber\\
	v^{\dagger}\!&:& {\mathbb{C}_\chi}^{\otimes 5} \rightarrow {\mathbb{C}_{\tilde{\chi}}}^{\otimes 2}, ~~~~~~~~~~~~ v^{\dagger} v = I_{\tilde{\chi}^2},\nonumber\\
  w^{\dagger}\!&:& {\mathbb{C}_\chi} \otimes {\mathbb{C}_{\tilde{\chi}}}^{\otimes 6} \rightarrow {\mathbb{C}_{\chi}}, ~~~~~~~w^{\dagger} w = I_{\chi}, \label{eq:tensors2}
\end{eqnarray}
transform the state $\ket{\Psi_\tau} \in {\mathbb{C}_\chi}^{\otimes N}$ of the lattice $\mathcal{L}_\tau$ into a state $\ket{\Psi_{\tau+1}}\in {\mathbb{C}_{\chi}}^{\otimes N/16}$ of the effective lattice $\mathcal{L}_{\tau+1}$.

\begin{figure}[!htb]
\begin{center}
\includegraphics[width=8cm]{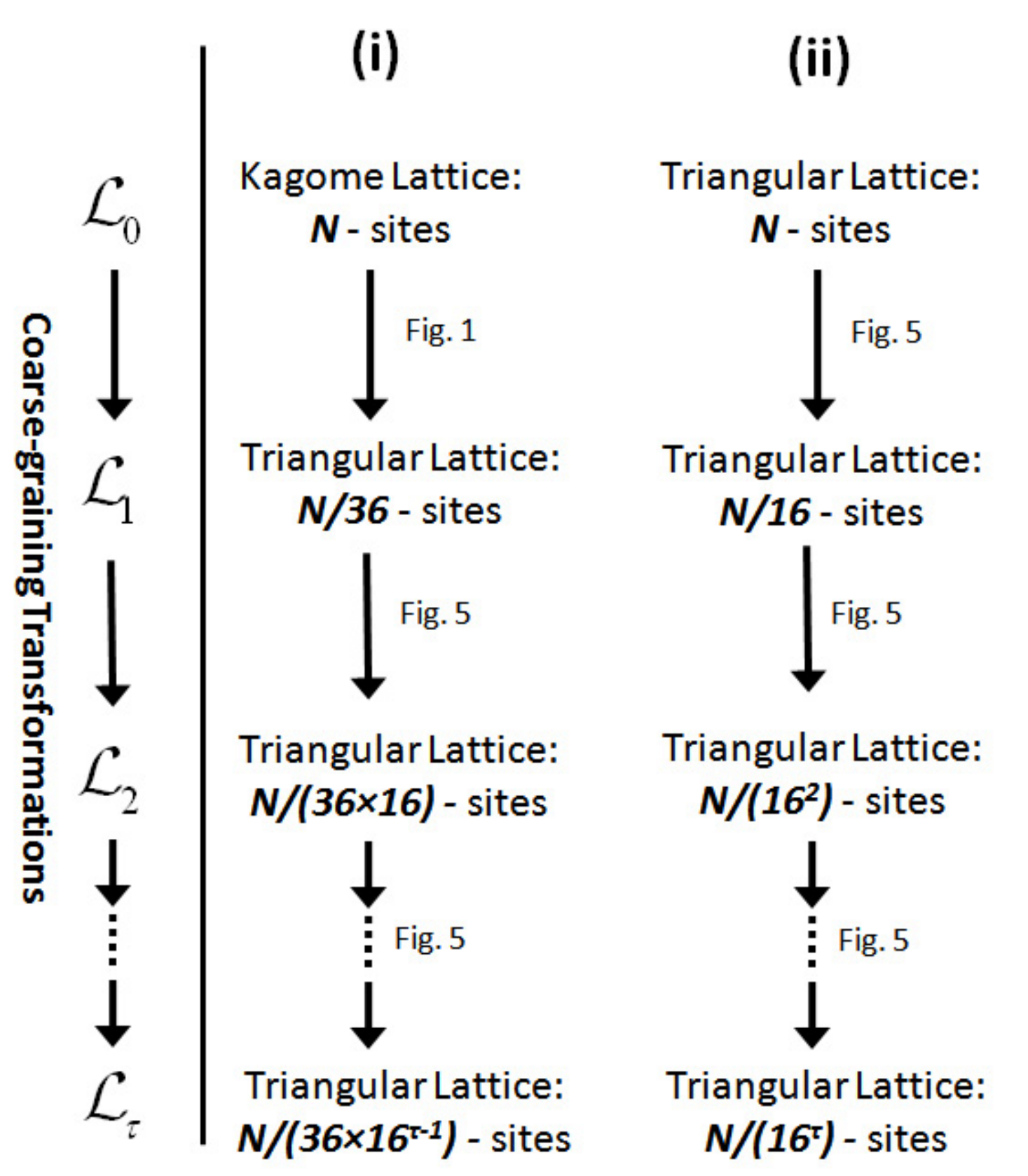}
\caption{(i) A kagome lattice MERA can be built using the transformation of Fig. \ref{fig:KagScheme} as the bottom MERA layer, equivalently as the initial coarse-graining transformation, and the transformation of Fig. \ref{fig:TriScheme} for subsequent MERA layers. (ii) The transformation of Fig. \ref{fig:TriScheme} can also be used directly to investigate models defined on triangular lattice geometries.}\label{fig:KagTri} 
\end{center}
\end{figure}

A MERA for the state $\ket{\Psi_0}$ of a kagome lattice system can be built by composing the transformation of Fig. \ref{fig:KagScheme} with several transformations as in Fig. \ref{fig:TriScheme}. Following this recipe, c.f. Fig. \ref{fig:KagTri}, a MERA of $\tau$ layers allows for investigations of finite kagome lattices of $N = 36\times 16^{\tau-1}$ sites. This scheme can also be used with the scale-invariant algorithm (Sect V.C of \cite{NewAlgorithm}) to investigate scale-invariant systems, such as critical systems, on infinite kagome lattices. The triangular coarse-graining transformation of Fig. \ref{fig:TriScheme} may also be used directly to investigate models defined on triangular lattices. A MERA with $\tau$ layers, each defined by the transformation Fig. \ref{fig:TriScheme}, represents a wave-function on a triangular lattice of $N = 16^\tau$ sites. Indeed, when applied to a spin-$\frac{1}{2}$ Heisenberg model on an $N=\infty$ triangular lattice, this scheme reproduces the three-sublattice $120^ \circ$ order found in other studies and gives a ground energy for the infinite system at $E = -0.545$ per site, comparable to a series expansion estimate of $E = -0.5502$ per site \cite{TriHeis}. \newline

%%%%%%%%%%%%%%%%%%%%%%%%%%%%%%%%%%%%%%%
%%%%%%%%%%%%%%%%%%%%%%%%%%%%%%%%%%%%%%%
%%%%%%%%%%%%%%%%%%%%%%%%%%%%%%%%%%%%%%%
%%%%%%%%%%%%%%%%%%%%%%%%%%%%%%%%%%%%%%%

%%%%%%%%%%%%%%%%%%%%%%%%%%%%%%%%%%%%%%%
%%%%%%%%%%%%%%%%%%%%%%%%%%%%%%%%%%%%%%%
\begin{figure}[!tb]
\begin{center}
\includegraphics[width=6cm]{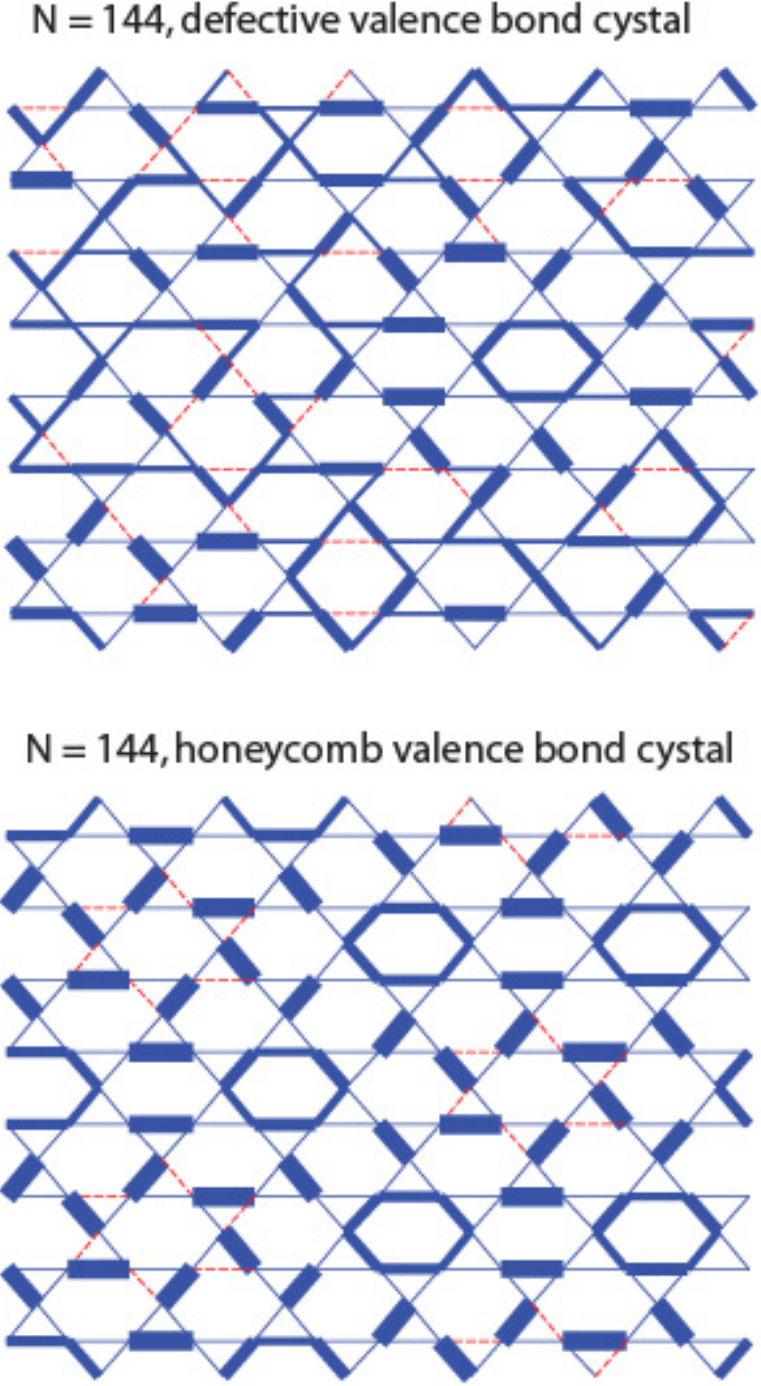}
\caption{(Color on-line) Converged $\tilde\chi=8$ MERA solutions for the $N=144$ site KLHM. The thickness of a line between two lattice sites is proportional to the absolute bond energy, $|E|$. Blue solid lines are used for negative energy bonds, $E<0$, and red dashed lines for positive energy bonds, $E>0$. (Top) An example of a `defective' VBC. Local structures that match the `pinwheels' and `perfect hexagons' of the honeycomb VBC are observed, but long range order is absent. (Bottom) By restricting the rate of lattice `cooling', as per Eq.\ref{eq:cooling}, one more reliably obtains the honeycomb VBC from a randomly initialized MERA. On average a defective VBC had $0.05\%$ greater energy than the honeycomb VBC of the same value of $\chi$. } \label{fig:BondPic}
\end{center}
\end{figure}
%%%%%%%%%%%%%%%%%%%%%%%%%%%%%%%%%%%%%%%
%%%%%%%%%%%%%%%%%%%%%%%%%%%%%%%%%%%%%%%

\textbf{Appendix B: Defective Crystals}

This appendix details a modification of the MERA optimization algorithm \cite{NewAlgorithm} that was found necessary to ensure proper convergence of some simulations of the KLHM.

All simulations of the KLHM converged to highly dimerized VBC type states. For lattices of $N=\{144,\infty\}$ sites the lowest energy state that was found, or best approximation to the ground-state, had a pattern of strong bonds matching the `honeycomb' VBC state. However, not all randomly initialized simulations converged to this state; in some instances the MERA wave-function converged to a `defective' VBC state, an example of which is shown Fig. \ref{fig:BondPic}. The observed defective VBC's ranged from states which matched the honeycomb VBC with only a few misplaced strong bonds to states with an almost completely disordered placement of strong bonds. A defective VBC was observed to have on average $0.05\%$ higher energy per site than the corresponding honeycomb VBC type MERA of the same $\tilde \chi$. The exact difference in energy depended on the amount of defects; typically, states that deviated more from the honeycomb VBC had higher energy than states which deviated less. In the instances that the method did not converge to the honeycomb VBC one may conclude that the optimization of the MERA became trapped in a local minimum.

Becoming trapped in a local minimum is not surprising given that the optimization of the MERA is based upon updates of individual tensors. During the initial iterations of the optimization, locally stable structures, such as singlets between neighboring spins, form. But they do not necessarily form in places compatible with a global minimization of the energy. Once the method has converged to a particular VBC state, the transformation required to bring the state to a different VBC, with a new pattern of strong bonds, requires simultaneous shifts of many strong bonds. This is unlikely to occur with an optimization based on energy-lowering updates of individual tensors.

There are many ways to decrease the risk of becoming trapped in a local minimum due to the formation of local singlets. In the present work this is achieved by restricting how much a tensor is allowed to change in one single update; this may be thought of as decreasing the rate at which the MERA is `cooled-down' from an initial (high-energy) state to the (low-energy) approximation to the ground-state. The goal of this restriction is to prevent locally stable structures from forming too early in the optimization. In order to discuss how this restriction may be implemented, we first refresh some details of the variational MERA optimization, see Sect. IV of \cite{NewAlgorithm}. In the standard algorithm, a tensor $w$ in the MERA is replaced with a better (that is leading to lower energy) tensor $w'$ as follows. First one computes the linearized environment of $w$, denoted $\Upsilon _w$. Then, given the singular value decomposition of the environment, $\Upsilon _w = U S V^\dag$, one chooses the updated tensor to be $w'=V U^\dag$. Here we wish to restrict the amount $\varepsilon$ the tensor can change, $\left\| {w - w'} \right\| < \varepsilon$. This is achieved by using a modified environment $\tilde \Upsilon _w$ defined as 
\begin{equation}
\tilde \Upsilon _w  \equiv \Upsilon _w  + \lambda w 
\label{eq:cooling}
\end{equation}
for some $\lambda>0$. As before, given the singular value decomposition of the modified environment $\tilde \Upsilon _w = U S V^\dag$, the updated tensor $w'$ is chosen to be $w'=V U^\dag$. The parameter $\lambda$ acts as a soft constraint to ensure the tensor change $\varepsilon$ is kept small with each update, a larger value of $\lambda$ giving a smaller change $\varepsilon$. In the limit of $\lambda\rightarrow\infty$, the tensor $w$ would remain constant, $\varepsilon=0$, in the update. Simulations of the KLHM with a suitable value of $\lambda$, say $\lambda>10$ for the initial optimization, were much more likely to converge to a defect free honeycomb VBC than were unrestrained ($\lambda=0$) optimizations.

\end{document}